\documentclass[aps,prl,amsmath,amssymb,reprint,superscriptaddress]{revtex4-1}

\usepackage{graphicx}
\usepackage{mathptmx,times}
\usepackage{xcolor}

\usepackage{CJK}

\usepackage[breaklinks=true,colorlinks=true,bookmarks=false,urlcolor=blue,
citecolor=blue,linkcolor=blue]{hyperref}

\def\unit#1{\mathord{\thinspace\rm #1}}

\begin{document}


\title{Anomalous Cyclotron Motion in Graphene Superlattice Cavities}

\author{Rainer Kraft}
\email{rainer.kraft@kit.edu}
\affiliation{Institute of Nanotechnology, Karlsruhe Institute of Technology, Karlsruhe D-76021, Germany}
\affiliation{Institute of Physics, Karlsruhe Institute of Technology, Karlsruhe D-76049, Germany}

\author{Ming-Hao Liu} 
\email{minghao.liu@phys.ncku.edu.tw}
\affiliation{Department of Physics, National Cheng Kung University, Tainan 70101, Taiwan}

\author{Pranauv Balaji Selvasundaram}
\affiliation{Institute of Nanotechnology, Karlsruhe Institute of Technology, Karlsruhe D-76021, Germany}
\affiliation{Department of Materials and Earth Sciences, Technical University Darmstadt, Darmstadt D-64287, Germany}

\author{Szu-Chao Chen} 
\affiliation{Department of Physics, National Cheng Kung University, Tainan 70101, Taiwan}

\author{Ralph Krupke}
\affiliation{Institute of Nanotechnology, Karlsruhe Institute of Technology, Karlsruhe D-76021, Germany}
\affiliation{Department of Materials and Earth Sciences, Technical University Darmstadt, Darmstadt D-64287, Germany}
\affiliation{Institute for Quantum Materials and Technologies, Karlsruhe Institute of Technology, Karlsruhe D-76021, Germany}

\author{Klaus Richter}
\affiliation{Institut f\"{u}r Theoretische Physik, Universit\"{a}t Regensburg, Regensburg D-93040, Germany}

\author{Romain Danneau}
\email{romain.danneau@kit.edu}
\affiliation{Institute of Nanotechnology, Karlsruhe Institute of Technology, Karlsruhe D-76021, Germany}
\affiliation{Institute for Quantum Materials and Technologies, Karlsruhe Institute of Technology, Karlsruhe D-76021, Germany}

\begin{abstract}

We consider graphene superlattice miniband fermions probed by electronic interferometry in magneto-transport experiments. By
decoding the observed Fabry-P\'{e}rot interference patterns together with our corresponding quantum transport simulations,
we find that the Dirac quasiparticles originating from the superlattice minibands do not undergo conventional cyclotron motion
but follow more subtle trajectories. In particular, dynamics at low magnetic fields is characterized by peculiar, straight trajectory
segments. Our results provide new insights into superlattice
miniband fermions and open up novel possibilities to use periodic potentials in electron optics experiments.

\end{abstract}

\maketitle


The presence of a superlattice potential modulates the intrinsic electronic band structure of a material \cite{Esaki1970}, allowing the observation of various physical phenomena such as Wannier-Stark ladders \cite{Mendez1988} or Weiss and Bloch oscillations \cite{Weiss1989,Waschke1993}. In graphene, superimposed long-range periodic potentials are predicted to alter the electronic dispersion with the emergence of extra singularities and new effective Dirac fermions \cite{Park2008prl,Park2008natphys,Barbier2010}. Thanks to the rapid development of artificial two-dimensional van der Waals (vdW) heterostructures, studying superlattice effects in graphene is nowadays possible \cite{Yankowitz2019}. For instance, when a graphene sheet is placed onto a hexagonal boron nitride (hBN) crystallite, interference due to the small lattice constant mismatch of about $1.8\,\%$ gives rise to a moir\'{e} pattern with a superstructure lattice parameter inversely proportional to the rotational misalignment between the layers. Strikingly, while secondary Dirac points appear in the modulated electronic band structure \cite{Yankowitz2012}, magneto-transport measurements revealed the Hofstadter butterfly \cite{Ponomarenko2013,Dean2013,Hunt2013} as well as magnetic Bloch states via the observation of Brown-Zak oscillations \cite{Ponomarenko2013,KrishnaKumar2017,Chen2017}. The latter highlights a particular metallic behavior with straight trajectories of the quasiparticles at relatively high magnetic field, \textit{i.\,e.}~at values of the magnetic flux per superlattice unit cell area commensurate with the magnetic flux quantum. At low magnetic field the impact of the superlattice on the quasiparticles has been studied by transverse electron focusing experiments showing that cyclotron motions break down near the secondary Dirac points \cite{Lee2016}. However, studying and distinguishing the transport behaviors of charge carriers due to superlattice minibands from those arising from the normal Dirac spectrum in the absence of magnetic field, as well as the crossover to the intermediate field regime, remains very challenging.

\begin{figure}
\includegraphics{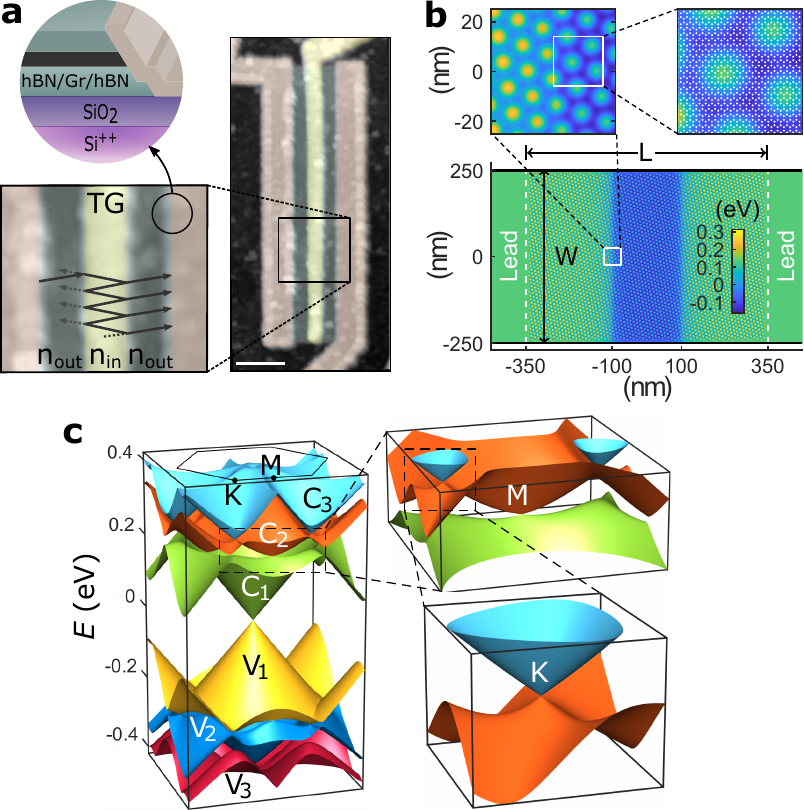}
\caption{(a)~False-color AFM image of the device (scale bar $1\unit{\mu m}$). The local top gate (TG, yellow) in the center forms in combination with the overall back gate (BG, green) a Fabry-P\'{e}rot cavity. Partially reflected and transmitted ballistic trajectories in this electronic interferometer are indicated by small arrows. The magnified region of the contact interface depicts a schematic of the edge-connected hBN/graphene/hBN vdW heterostructure. (b)~Schematic of the modeled device, showing the (scaled) graphene lattice as the scattering region with a superimposed periodic long-range potential. (c)~Corresponding electronic band structure of a single-layer graphene with electrostatic superlattice based on the continuum model. Close-ups of the minibands at special points $M$ and $K$ of the superlattice mini-Brillouin zone are shown on the right.}
	\label{fig:1}
\end{figure}

\begin{figure*}
\includegraphics[width=0.98\textwidth]{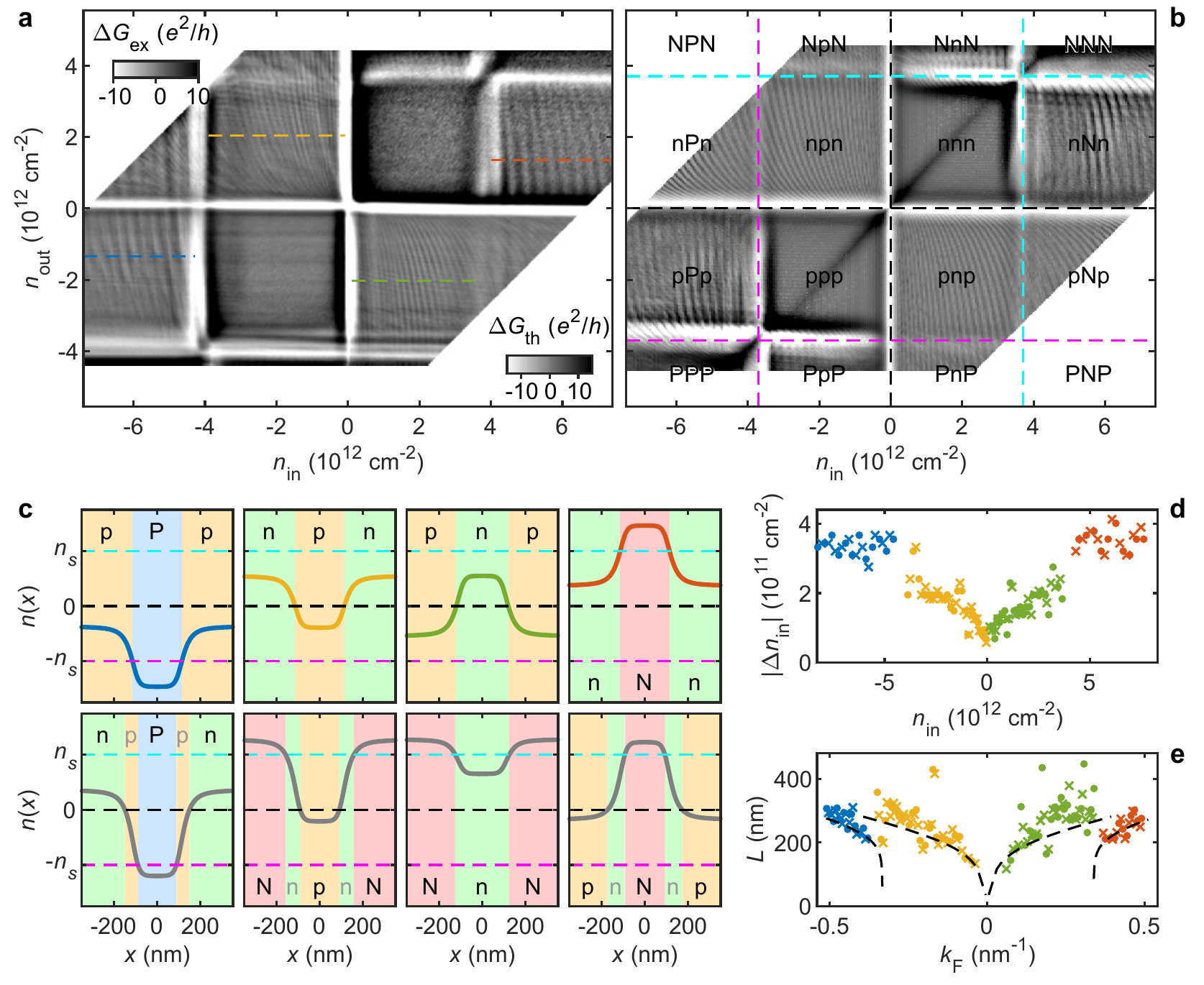}
\caption{(a,\,b)~Measured and simulated map of the conductance with subtracted smooth background, respectively, as a function of $n_\mathrm{in}$ and $n_\mathrm{out}$. Labels denote the charge carrier configurations in all $16$ quadrants, which are defined by the main Dirac point (black dashed lines) and hole/electron satellite Dirac point (magenta/cyan dashed lines). (c)~Spatial density profiles across the device for different doping scenarios. (d)~Density spacing between resonances as a function of $n_\mathrm{in}$ for the cavities pPp (blue), npn (orange), pnp (green) and nNn (red) at constant outer charge carrier densities $n_\mathrm{out}=-1.4\times10^{12}\,$cm$^{-2}$, $2.0\times10^{12}\,$cm$^{-2}$, $-2.0\times10^{12}\,$cm$^{-2}$ and $1.4\times 10^{12}\,$cm$^{-2}$, respectively, as marked by the colored dashed lines in panel (a). (e)~Corresponding cavity size $L$ as a function of Fermi wave vector $k_\mathrm{F}$. The black dashed lines show a numerical determination of the cavity size obtained from electrostatic simulations.}
\label{fig:2}
\end{figure*}

Here, we report a study of superlattice Dirac fermion transport through an electrostatically defined cavity formed by a local top gate (TG) and an overall back gate (BG) (see Fig.~\ref{fig:1}a). The combination of the two gates allows independent control over the charge carrier densities $n_\mathrm{in}$ and $n_\mathrm{out}$ in inner and outer regions of the device, respectively. Due to semi-transparent boundaries in a bipolar configuration (\textit{i.\,e.}~transitions across the charge neutrality point between inner and outer regions) a Fabry-P\'{e}rot (FP) cavity forms, where the interference of partially reflected and transmitted ballistic charge carrier trajectories gives rise to conductance oscillations \cite{Young2009,Campos2012,Varlet2014,Du2018}. In the case of a moir\'{e} superlattice the transitions across secondary Dirac points result in the formation of extra cavities \cite{Handschin2017}. By investigating the unusual cavity properties and probing the arising interferences via magneto-transport measurements, as well as quantum transport simulations, we are able to study the moir\'{e} miniband conduction associated with unconventional Dirac quasiparticle dynamics.
Figures~\ref{fig:1}b,\,c display a schematic of the simulated device geometry with superimposed periodic long-range potential as the scattering region \cite{Liu2015,Chen2019}, and the corresponding electronic band structure obtained from the continuum model, respectively (details given in Ref.~\onlinecite{Chen2019}). 

The effect of the superlattice in graphene is directly observed in transport experiments and a moir\'{e} wavelength $\lambda\approx10.9\,$nm has been extracted from the analysis of the Brown-Zak oscillations (see Supplemental Materials for more details \cite{supmat}). As overview on the density-dependent device characteristics, Figs.~\ref{fig:2}a,\,b show maps of the conductance as a function of $n_\mathrm{in}$ and $n_\mathrm{out}$ obtained from experimental measurement and quantum transport simulation, respectively (see Supplemental Material \cite{supmat} for raw resistance and conductance as a function of applied BG and TG voltage). The conductance profile is structured by the appearance of the main Dirac point at vanishing densities (highlighted by black dashed lines in Fig.~\ref{fig:2}b) and satellite Dirac points on electron- and hole-side (highlighted by cyan and magenta dashed lines in Fig.~\ref{fig:2}b), resulting in a map composed of 16 regions of unique doping configurations. To facilitate the following discussions, we introduce a notation with small letters p and n describing charge carriers within the linear valence band V$_1$ and conduction band C$_1$ of the original primary Dirac cone, while capital letters P and N denote charge carriers in the back-folded superlattice minibands (V$_2$, V$_3$ and C$_2$, C$_3$) below and above the secondary Dirac points, respectively. The labeling of each junction configuration is shown in Fig.~\ref{fig:2}b. We note that in the experimental map the two regions on the upper left corner (NPN) and lower right corner (PNP) are missing, corresponding to the most extreme opposite doping between inner and outer regions where applied TG and BG voltages counteract in the dual-gated part of the device.

In both maps we observe multiple sets of FP interferences originating from the different cavity combinations. Notably, in addition to the well-known resonances in the bipolar regions npn and pnp \cite{Young2009,Campos2012,Varlet2014,Du2018}, interference patterns are visible in the nominal unipolar quadrants of an ordinary graphene p--n junction device. Here, cavities can be formed where the interfaces between regions of charge carriers from the normal Dirac spectrum and superlattice minibands play the role of semi-transparent boundaries rather than p--n interfaces (cf.~ref.~\onlinecite{Handschin2017}). Indeed, the co-existence of both junction types can be strikingly observed in Fig.~\ref{fig:2}a with two superimposed but distinct sets of FP interferences in nPn (pNp). For these configurations, the spatial density profile from inner to outer regions features transitions first across the satellite Dirac point and then across the primary Dirac point (see Fig.~\ref{fig:2}c). Hence, the inner most cavity is formed in the same fashion as in pPp (nNn) and consequently, oscillation fringes originating in pPp (nNn) remain visible across the main Dirac peak of the outer charge carrier density axis. We note that in the simulated map (Fig.~\ref{fig:2}b) the continuation of the fringes is only faintly visible in pNp.

We now compare the behavior of a ``common'' p--n--p junction with a cavity formed by the satellite Dirac points, focusing on configurations npn, pnp and pPp, nNn (see Fig.~\ref{fig:2}c for corresponding spatial density profiles). Though it should be mentioned that the moir\'{e} superlattice potential is always present in both inner and outer regions, only the Fermi level is spatially tuned by the BG and TG to reside within the different bands of the reconstructed band structure (unlike the transition from a non-superlattice to superlattice region as \textit{e.\,g.}~proposed in ref.~\onlinecite{Beenakker2018}). It is already visible from the maps (Figs.~\ref{fig:2}a,\,b) that the spacing of the resonances in pPp and nNn appears notably increased in comparison to cavities npn and pnp. In Fig.~\ref{fig:2}d the extracted density spacing is plotted as a function of $n_\mathrm{in}$ for exemplary linecuts at constant $n_\mathrm{out}$ (also see Supplemental Material for corresponding conductance curves \cite{supmat}), highlighting the observed trend of increased spacing (likewise \textit{cf.}~ref.~\onlinecite{Handschin2017}). The enlarged oscillation period can be attributed to a suddenly reduced cavity size due to the newly defined boundaries by the satellite Dirac points (see Supplemental Material \cite{supmat}). Figure~\ref{fig:2}e shows the observed drop in the experimentally extracted cavity size $L=\pi /\Delta k_\mathrm{F}\,$, where $\Delta k_\mathrm{F} = \left | k_\mathrm{F,\,j+1} - k_\mathrm{F,\,j} \right |$ with $k_\mathrm{F}=\sqrt{\pi n}$ and $j$ the interference maximum index, following in good agreement the trend of a numerical determination of the cavity size from electrostatic simulations (black dashed lines).

\begin{figure*}
\includegraphics{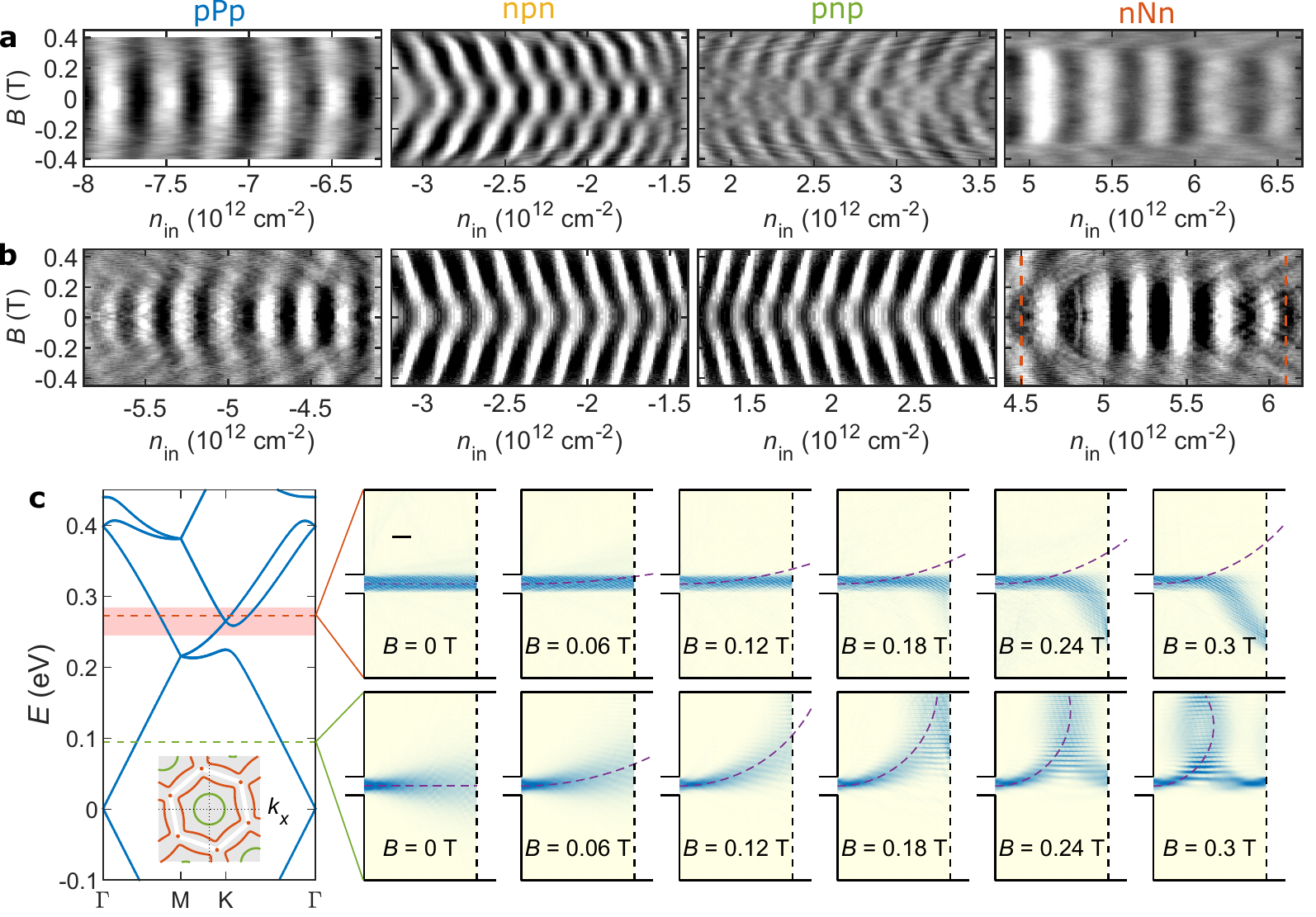}
\caption{Low magnetic field measurements of the four different cavities pPp, npn, pnp and nNn in column-wise order at constant outer charge carrier densities $n_\mathrm{out}=-1.4\times10^{12}\,$cm$^{-2}$, $2.0\times10^{12}\,$cm$^{-2}$, $-2.0\times10^{12}\,$cm$^{-2}$ and $1.4\times10^{12}\,$cm$^{-2}$, respectively. (a,\,b)~Experimental and simulated patterns of the conductance oscillations, respectively, as a function of charge carrier density $n_\mathrm{in}$ and magnetic field $B$. (c)~Band structure of our superlattice model, with the red shaded region denoting the energy window of non-dispersing fringes in nNn. The inset depicts the Fermi surface slightly above the $K$ point of the mini-Brillouin zone marked by the red dashed line. Right-hand side panels: charge carrier beams are shown from simulations for varying magnetic field $B$ at two different Fermi levels at distinctly different band structure regions. Upper row: C$_2$ and C$_3$ minibands (red dashed line) Lower row: C$_1$ band of the primary Dirac cone (green dashed line).}
\label{fig:4}
\end{figure*}

To further explore the transport behavior of the charge carriers in the superlattice minibands, we have studied the four different cavities with applied magnetic field $B$ (as well as by source-drain bias spectroscopy, shown in the Supplemental Material \cite{supmat}). Figures~\ref{fig:4}a,\,b show maps of the measured and simulated FP interference patterns as a function of $n_\mathrm{in}$ and magnetic field $B$ at constant $n_\mathrm{out}$ (see Supplemental Material for additional patterns at different $n_\mathrm{out}$ \cite{supmat}). The resulting patterns in npn and pnp of both simulation and experiment show fringes with typical dispersing behavior towards higher densities with increasing magnetic field due to the competition of Aharonov-Bohm phase and Wentzel-Kramer-Brillouin kinetic phase, as well as the appearance of a $\pi$-shift as the hallmark of Klein tunneling in monolayer graphene \cite{Shytov2008,Young2009,Katsnelson,Rickhaus2015}. However, strikingly different patterns are observed in the case of pPp and nNn. Here, no $\pi$-shift is visible and fringes do not or only weakly disperse under the influence of the magnetic field which implies a lacking of a magnetic-field-dependent phase. Moreover, the resonances in nNn vanish abruptly at a critical field $B_\mathrm{c}\sim 300\,$mT, while in pPp the oscillation fringes remain visible, yet less pronounced. Nevertheless, the non-dispersing fringes are monitored in both cases.

In order to understand why the miniband charge carriers do not pick up a magnetic phase, we have investigated their trajectories by performing quantum transport simulations using a narrow beam injector into a wider 2D sample (see Supplemental Material for details \cite{supmat}). Figure~\ref{fig:4}c depicts the band structure with the red shaded region marking the shown energy window of non-dispersing fringes in nNn. The resulting trajectory beams for two distinct Fermi levels (red and green dashed lines) are shown on the right-hand side (for a complete set of beam simulations under various conditions see Supplemental Material \cite{supmat}). In the case of normal Dirac fermions originating from the primary Dirac cone (lower row) the beam follows the expected cyclotron motion of moving charge carriers in a magnetic field (the purple dashed lines correspond to the calculated cyclotron radius $r_\mathrm{c}=\hbar k_\mathrm{F} /eB$). In the other case (upper row), the simulated trajectories of the miniband charge carriers persist on a quasi-straight beam line on the length scale of the cavity for small magnetic fields, which explains the absence of the magnetic field dependent phase in the FP interference patterns.

The observed unusual magnetic field independence of the beam can be understood as the consequence of a reshaped Fermi surface of the superlattice minibands, shown as inset in Fig.~\ref{fig:4}c (left panel) for the two above discussed scenarios (green and red indicated Fermi levels, respectively). Following from the semiclassical equation of motion $\hbar\mathbf{\dot{k}}=q(\mathbf{E}+\mathbf{\dot{r}}\times\mathbf{B})$ (for a full description, see chapter 12 of \cite{AshcroftMermin}) the cyclotron orbit in real space is given by the orbit in momentum space rotated by $90\,^\circ$. Indeed, at increasing magnetic field a hexagonal bending of the trajectories can be noticed in opposite rotation direction, directly reflecting the hole-type hexagonal Fermi surface of C$_2$ (see Supplemental Material for further details \cite{supmat}). In fact, this hexagonal-like beam deflection on a length scale shorter than the cavity at higher magnetic fields causes the observed breakdown of FP interferences at $B\sim0.3\,$mT due to the oblique incident angle of charge carriers onto the cavity barriers (see Supplemental Material \cite{supmat}). In this regard, it is also important to note that the relative device orientation with respect to the superlattice structure plays an important role unlike the case of a circular Fermi surface in unperturbed graphene.

Finally, we observed different magnetic field dependence of FP interferences in NpN and NnN (see Supplemental Material \cite{supmat}). Interestingly, in both cases it is normal Dirac fermions confined to the cavity. Yet, the same argument of the reshaped Fermi surface holds to explain the distinct behaviors. In the case of NnN, the almost normal incident angle of superlattice quasiparticles from the outer reservoirs onto the cavity interfaces prevents the confined electrons to form closed loops inside the cavity (see Supplemental Material for details \cite{supmat}). At a magnetic field value, where the hexagonal orbital bending of the outside miniband fermions occurs on a length scale shorter than the outer reservoirs, the resonances vanish similar to the reversed configuration nNn. On the other hand, NpN features transitions across both primary and secondary Dirac points (see Fig.~\ref{fig:2}c). Since the innermost cavity resembles npn, a continuation of interference fringes from npn to NpN in the maps of Figs.~\ref{fig:2}a(b) can be observed, and the magnetic field behavior of the conductance resonances conforms with ordinary n--p--n junctions.

To conclude, we have probed the transport properties of charge carriers in a graphene/hBN moir\'{e} superlattice device by electronic interferometry. 
The non-dispersing fringes of FP interferences at low magnetic field combined with our quantum transport simulations revealed unconventional cyclotron motion on the length scale of the cavity which reflect the reshaped hexagonal Fermi surface.
The subtle transport properties of these quasiparticles provide a new versatile platform for new types of devices in electron optics experiments, which could be used to probe, for example, correlated states in twisted bilayers \cite{Cao2018,Cao2018a,Sharpe2019,Lu2019,Shen2020}.

\acknowledgments{The authors thank \ifmmode \dot{I}\else \.{I}\fi{}.~Adagideli, I.~Gornyi, A.~Knothe, P.~Makk, C.~Sch\"{o}nenberger and O.~Zilberberg for fruitful discussions. This work was supported by the Helmholtz society through the STN program, the German Research foundation (DFG, Deutsche Forschungsgemeinschaft) Project-ID 314695032-CRC1277 and through the Project Ri681/14-1, as well as the Taiwan Ministry of Science (107-2112-M-006-004-MY3 and 107-2627-E-006-001) and Ministry of Education (Higher Education Sprout Project).

\end{document}


\title{Supplemental Material for \\``Anomalous Cyclotron Motion in Graphene Superlattice Cavities''}

\author{Rainer Kraft}
\email[{\color{blue}rainer.kraft@kit.edu}]{}
\affiliation{Institute of Nanotechnology, Karlsruhe Institute of Technology, Karlsruhe D-76021, Germany}
\affiliation{Institute of Physics, Karlsruhe Institute of Technology, Karlsruhe D-76049, Germany}

\author{Ming-Hao Liu}
\email[{\color{blue}minghao.liu@phys.ncku.edu.tw}]{}
\affiliation{Department of Physics, National Cheng Kung University, Tainan 70101, Taiwan}

\author{Pranauv Balaji Selvasundaram}
\affiliation{Institute of Nanotechnology, Karlsruhe Institute of Technology, Karlsruhe D-76021, Germany}
\affiliation{Department of Materials and Earth Sciences, Technical University Darmstadt, Darmstadt D-64287, Germany}

\author{Szu-Chao Chen}
\affiliation{Department of Physics, National Cheng Kung University, Tainan 70101, Taiwan}

\author{Ralph Krupke}
\affiliation{Institute of Nanotechnology, Karlsruhe Institute of Technology, Karlsruhe D-76021, Germany}
\affiliation{Department of Materials and Earth Sciences, Technical University Darmstadt, Darmstadt D-64287, Germany}
\affiliation{Institute for Quantum Materials and Technologies, Karlsruhe Institute of Technology, Karlsruhe D-76021, Germany}

\author{Klaus Richter}
\affiliation{Institut f\"{u}r Theoretische Physik, Universit\"{a}t Regensburg, Regensburg D-93040, Germany}

\author{Romain Danneau}
\email[{\color{blue}romain.danneau@kit.edu}]{}
\affiliation{Institute of Nanotechnology, Karlsruhe Institute of Technology, Karlsruhe D-76021, Germany}
\affiliation{Institute for Quantum Materials and Technologies, Karlsruhe Institute of Technology, Karlsruhe D-76021, Germany}

\maketitle
\clearpage

\section{Sample fabrication and experimental details}
The presented results are based on a graphene/hBN van der Waals heterostructure. Starting with mechanically exfoliated graphene flakes from natural bulk graphite (NGS Naturgraphit GmbH) and hBN flakes from commercial hBN powder (Momentive, grade PT110), the selected single-layer graphene is then encapsulated between two hBN multilayers by sequentially piling up the three different 2D crystallites (thicknesses of top and bottom hBN are about $40\,$nm and $20\,$nm respectively). The crystallographic orientation of the graphene is aligned under a misorientation angle of about $\approx0.9\,^\circ$ with respect to one of the hBN layers resulting in the formation of the moir\'{e} superlattice and the final stack is sitting on a Si/SiO$_2$ substrate which serves as BG. The stacking procedure is done via the polymer-free assembly technique introduced in ref.~\onlinecite{Wang13} with minor adaptions (similarly to \cite{Kraft18}).

The device is then designed with a narrow local Cr/Au TG electrode in the center of the device forming an electrostatically tunable pnp junction in combination with the overall BG. Electrical contact to the graphene is made from the edge of the mesa, where we use a single resist layer of PMMA for both etching of the graphene/hBN stack and metalization of (superconducting) Ti/Al electrodes. The self-aligned metal contacts due to utilizing the same already patterned resist for subsequent metal deposition ensure high quality electrical connection with low contact resistance. In a final step, the devices are etched into the desired shape. It needs to be said that the TG electrodes were designed and deposited first before the final etching. Thus, there is a remaining additional narrow graphene sleeve underneath the TG on the side of the device from which the electrode is launched. However, considering a wide and short junction, this issue should not conflict the presented results of this work.

The experiments were performed in a $^3$He/$^4$He dilution fridge BF-LD250 from BlueFors at a temperature of about $200$--$250\,$mK, unless otherwise mentioned. Electrical measurements were conducted in a two-terminal configuration, using standard low-frequency ({$\sim13\,$Hz}) lock-in technique with low ac excitation ($<\SI{10}{\micro\volt}$). For the electrostatic gating and source-drain biasing ultra-low noise dc-power supplies from Itest were used with an additional high power supply for the BG. All magnetic field measurements were performed in an out-of-plane magnetic field and $B=20\,$mT were applied for the measurements at temperatures below $1\,$K to suppress effects of the proximity-induced superconductivity due to the superconducting Ti/Al contacts.

\section{Uniform doping characteristics}

\begin{figure*}
	\centering
	\includegraphics{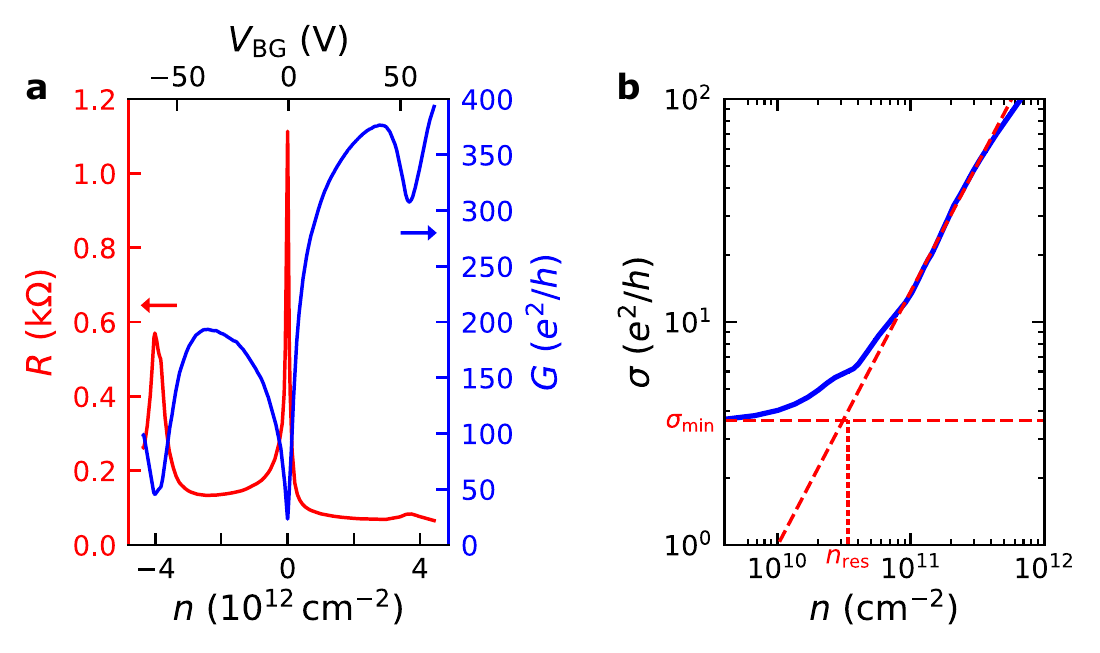}
	\caption{\setstretch{1.1} (a)~Resistance $R$ (red) and conductance $G$ (blue) as a function of back gate voltage $V_\mathrm{BG}$ (top axis) or converted overall charge carrier density $n=n_\mathrm{in} = n_\mathrm{out}$ (bottom axis) measured at $V_\mathrm{TG}=V_\mathrm{TG}^\mathrm{cnp}$. (b)~Conductivity $\sigma$ with subtracted contact resistances $2R_\mathrm{C}=57\,\Omega$ on the electron-side of the main Dirac point as a function of charge carrier density in a double logarithmic representation, giving an estimate for the residual charge carrier density.
	}
	\label{fig:S1}
\end{figure*}

The normal graphene field effect characteristics at uniform doping are shown in \ref{fig:S1}. At maximum charge carrier density $n\approx4.4\times10^{12}\,$cm$^{-2}$ the measured two-terminal resistance $R=66\,\Omega$ (see \ref{fig:S1}a). With the quantum resistance $R_\mathrm{Q}=h/(ge^2)\times(1/M)=9\,\Omega$ (where $M=W/(\lambda_\mathrm{F}/2)=710$ is the integer number of conductance modes and $g=4$ accounts for spin- and valley degeneracy) we find the contact resistance $R_\mathrm{C}=(R-R_\mathrm{Q})/2=28.5\,\Omega$ and resistivity $\rho_\mathrm{C}=171\,\Omega\mu$m. The residual charge carrier density is estimated \cite{Du08} on the electron side of the primary Dirac point as $n_\mathrm{res}\sim3\times10^{10}\,$cm$^{-2}$ (see \ref{fig:S1}b).

\begin{figure*}
	\centering
	\includegraphics[width=\textwidth]{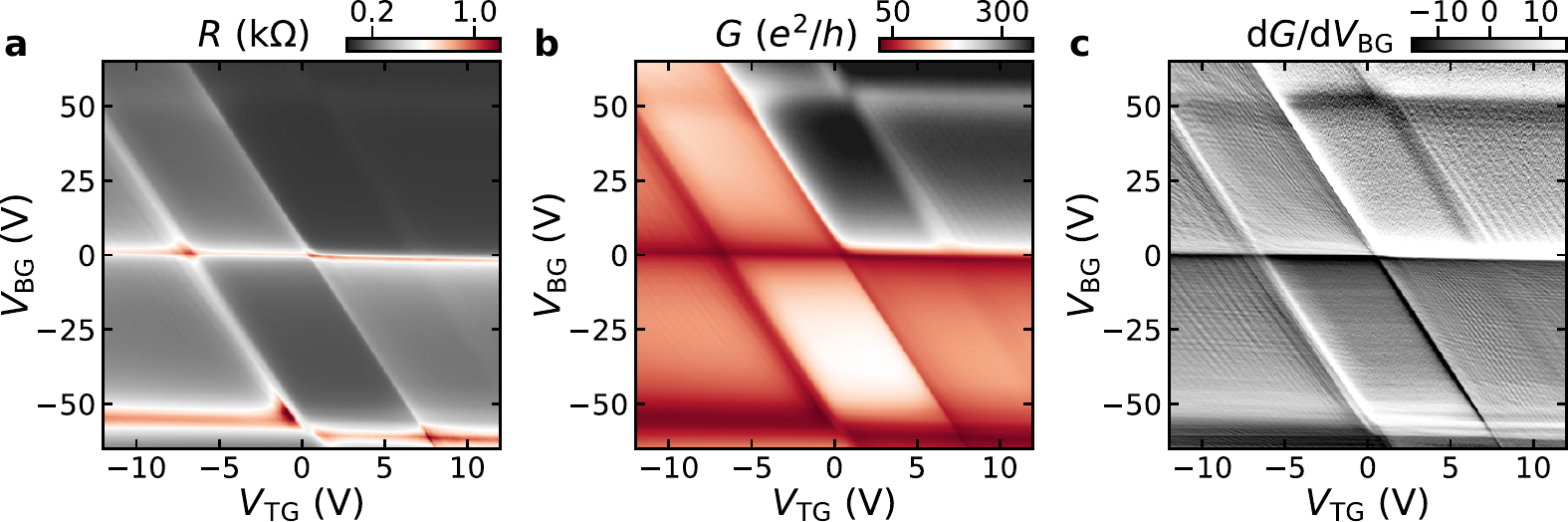}
	\caption{\setstretch{1.1} Additional gate-gate-map of the graphene moir\'{e} superlattice electronic interferometer device with raw two-terminal transport data, showing (a)~resistance $R$, (b)~conductance $G$ and (c)~differentiated conductance $\mathrm{d}G/\mathrm{d}V_\mathrm{BG}$ as a function of top gate voltage $V_\mathrm{TG}$ and back gate voltage $V_\mathrm{BG}$. The maps consist of sub-quadrants as discussed in the main text, but here defined by horizontal lines corresponding to charge neutrality of the outer regions (tuned by the back gate) and diagonal lines corresponding to charge neutrality of the dual-gated region (controlled by both top and back gate).}
	\label{fig:S5}
\end{figure*}

\begin{figure*}
	\centering
	\includegraphics{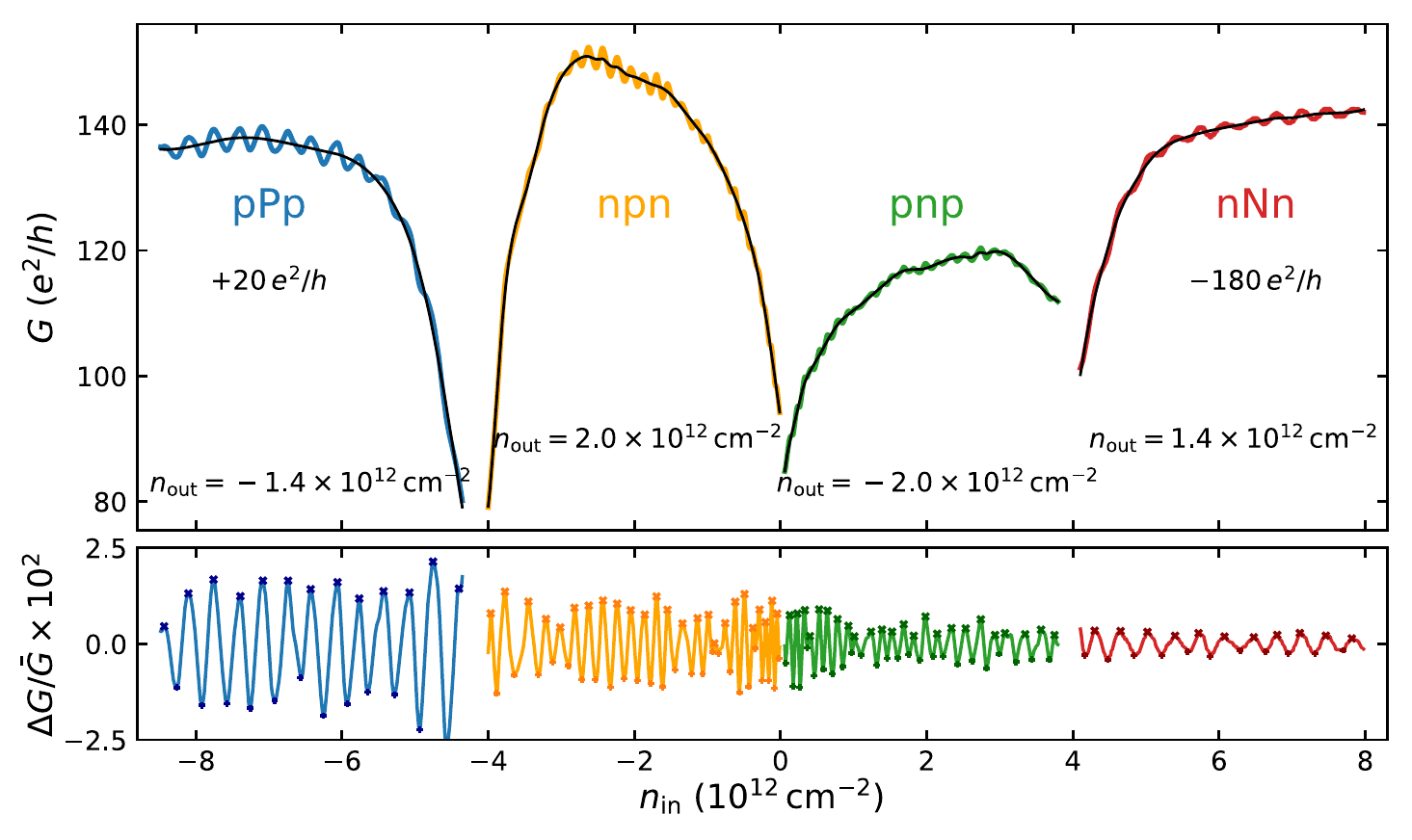}
	\caption{\setstretch{1.1} Top: Conductance $G$ as a function of charge carrier density $n_\mathrm{in}$ for cavities pPp (blue), npn (orange), pnp (green) and nNn (red) (corresponding to the data in Fig.~2 of the main text). Curves of pPp and nNn have been shifted by a constant offset (see numbers in the plot) for better comparison of all four cavities. Thinner black lines correspond to the smooth conductance background $\bar{G}$, which is subtracted to obtain the net conductance oscillations $\Delta G = G - \bar{G}$\,. Bottom: Normalized oscillation amplitude $\Delta G/\bar{G}$. Positions of maximums and minimums are marked by 'x' or '+', respectively, from which the spacings $\Delta n_\mathrm{in}$ were extracted (shown in the main text).}
	\label{fig:S6}
\end{figure*}

\section{Additional cavity analysis}

\ref{fig:S5} displays, in turns, the raw resistance $R$ (a), conductance $G$ (b) and differentiated conductance $\mathrm{d}G/\mathrm{d}V_\mathrm{BG}$ (c) versus back gate and top gate voltages $V_\mathrm{BG}$ and $V_\mathrm{TG}$. Charge carrier densities in the inner and outer regions of the device are converted from back gate and top gate voltages $V_\mathrm{BG}$ and $V_\mathrm{TG}$ as follows: $n_\mathrm{in}=\frac{C_\mathrm{BG}}{e}(V_\mathrm{BG}-V_\mathrm{BG}^\mathrm{cnp})+\frac{C_\mathrm{TG}}{e}(V_\mathrm{TG}-V_\mathrm{TG}^\mathrm{cnp})$ and $n_\mathrm{out}=\frac{C_\mathrm{BG}}{e}(V_\mathrm{BG}-V_\mathrm{BG}^\mathrm{cnp})$, where $C_\mathrm{BG}=0.67\times10^{11}\,$cm$^{-2}$ and $C_\mathrm{TG}=5.73\times10^{11}\,$cm$^{-2}$ are the specific gate capacitances per unit area and $V_\mathrm{BG}^\mathrm{cnp}=-0.1\,$V and $V_\mathrm{TG}^\mathrm{cnp}=0.45\,$V are offset voltages of the charge neutrality point, respectively. The specific BG capacitance $C_\mathrm{BG}$ is determined from the Landau level fan diagram as a function of $V_\mathrm{BG}$ for approximately uniform doping at $V_\mathrm{TG}=V_\mathrm{TG}^\mathrm{cnp}$. The specific SG capacitance $C_\mathrm{SG}$ is then extracted from the lever arm of tuning the charge neutrality point with respect to the back gate $C_\mathrm{SG}=C_\mathrm{BG}|\Delta V_\mathrm{BG}/\Delta V_\mathrm{SG}|$.

\begin{figure*}
\centering
\includegraphics[width=\textwidth]{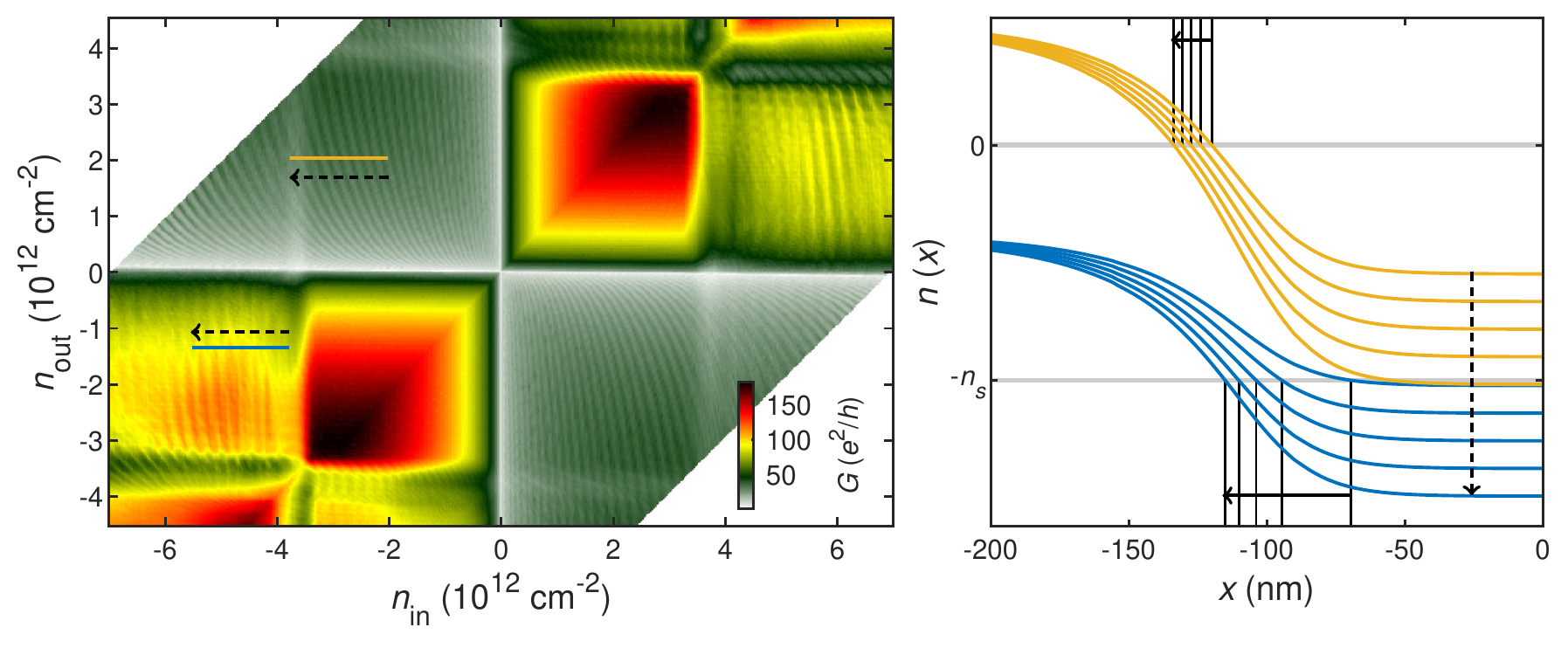}
\caption{\setstretch{1.1}Cavity size. Left: Conductance $G$ as a function of $n_\mathrm{in}$ and $n_\mathrm{out}$ from quantum transport simulations (cf.~Fig.~2b of the main text). Right: Charge carrier density profiles across the left $n_\mathrm{out}$--$n_\mathrm{in}$ cavity interface obtained from electrostatic simulations. The profiles denote respective inner and outer charge carrier density configurations npn (orange) and pPp (blue) as indicated by the colored lines in the left panel. From top to bottom corresponds to an increasing inner charge carrier density. The cavity size abruptly shrinks when $n_\mathrm{in}$ is tuned across the secondary Dirac point (see transition from lower most orange curve to upper most blue curve; also cf.~Fig.~2e of the main text).}
\label{fig:S7}
\end{figure*}

\ref{fig:S6} displays conductance oscillations in normal and superlattice Fabry-P\'{e}rot cavities for various carrier densities of the outer regions, corresponding to the Fig.2 of the main text. As demonstrated in \cite{Handschin17}, the size of the cavity strongly varies with the top gate voltage, in particular at low charge carrier density. \ref{fig:S7}a shows quantum transport simulations of the conductance $G$ versus $n_\mathrm{in}$ and $n_\mathrm{out}$. Two density profiles $n(x)$ are extracted from \ref{fig:S7}a and plotted on \ref{fig:S7}b for various $n_\mathrm{in}$, demonstrating that the cavity size clearly shrinks passing the secondary Dirac point.

\section{Additional magneto-transport: Landau level fan, Brown-Zak oscillations and Fabry-P\'{e}rot interferences}

\begin{figure*}
	\centering
	\includegraphics[width=\textwidth]{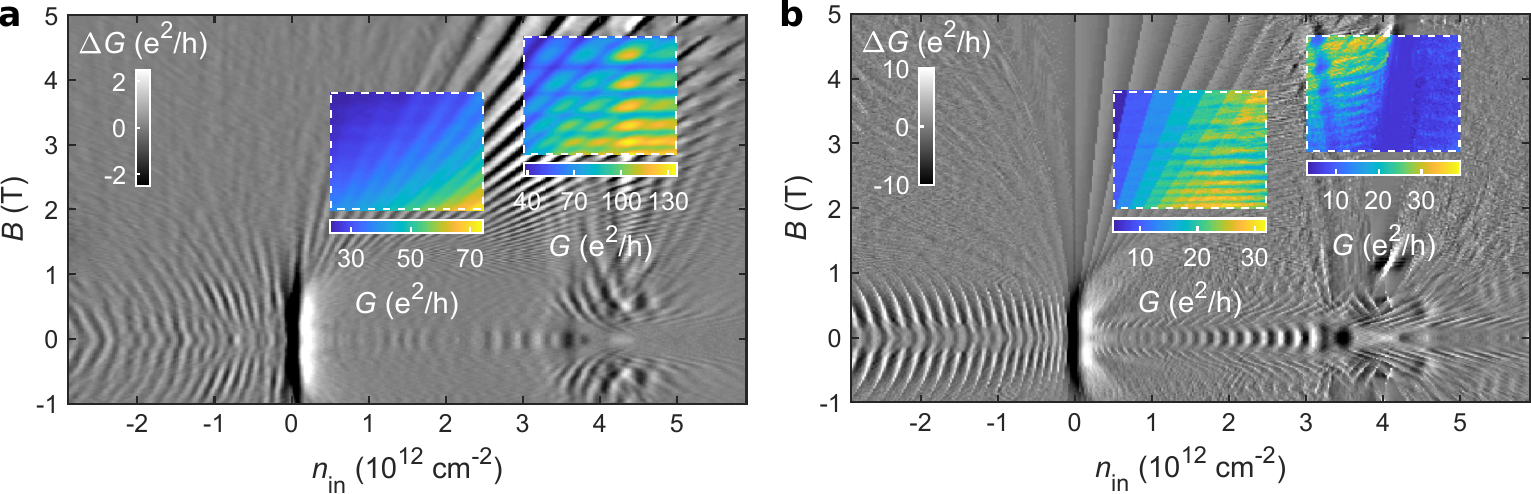}
	\caption{\setstretch{1.1} (a)~Measured and (b)~simulated conductance map with subtracted smooth background as a function of inner charge carrier density $n_\mathrm{in}$ and magnetic field $B$ for a constant outer charge carrier density $n_\mathrm{out}=4.1\times 10^{12}\,$cm$^{-2}$, \textit{i.\,e.}~the inner cavity is connected to reservoirs with charge carriers of the electron-side moir\'{e} minibands. In both panels colored inset plots show raw conductance.}
	\label{fig:S2}
\end{figure*}

Mapping the conductance at low and medium magnetic field shows several intriguing phenomena in high quality graphene devices. \ref{fig:S2} displays both experiment (a) and simulation (b) conductance maps, respectively, as a function of the charge carrier density $n_\mathrm{in}$ and magnetic field $B$ at constant outer charge carrier density $n_\mathrm{out}=4.1\times 10^{12}\,$cm$^{-2}$, \textit{i.\,e.}~above the electron-side secondary Dirac point in the outer reservoirs. The conductance is plotted with subtracted smooth background in the same way as \cite{Du18}. The features in the experimental map are well reproduced by our scalable tight-binding simulations \cite{Liu15} with the adapted quantum transport model for electrostatic superlattices \cite{Chen19} to our device geometry. At both primary and secondary Dirac points of the inner cavity emerging Landau level fans are observed, which are known to make up the Hofstadter butterfly spectrum at high magnetic fields \cite{Ponomarenko13,Dean13,Hunt13}. Furthermore, Brown-Zak oscillations are visible as horizontal lines in the spectrum \cite{Ponomarenko13,KrishnaKumar17,Chen17} (see colored inset panels with raw conductance in \ref{fig:S2}a and b). From the frequency of these oscillations we have determined the moir\'{e} wavelength of our superlattice structure (see below). Most notably, at low magnetic field, distinct unusual conductance oscillation patterns are observed in the junction configurations NpN and NnN, respectively (see discussion main text). The particular FP interference patterns reveal the strong sensitivity of transport through the induced electronic interferometer on the formed cavity.

\section{Estimation of the moir\'{e} wavelength of the superlattice structure}

\begin{figure*}
	\centering
	\includegraphics[width=\textwidth]{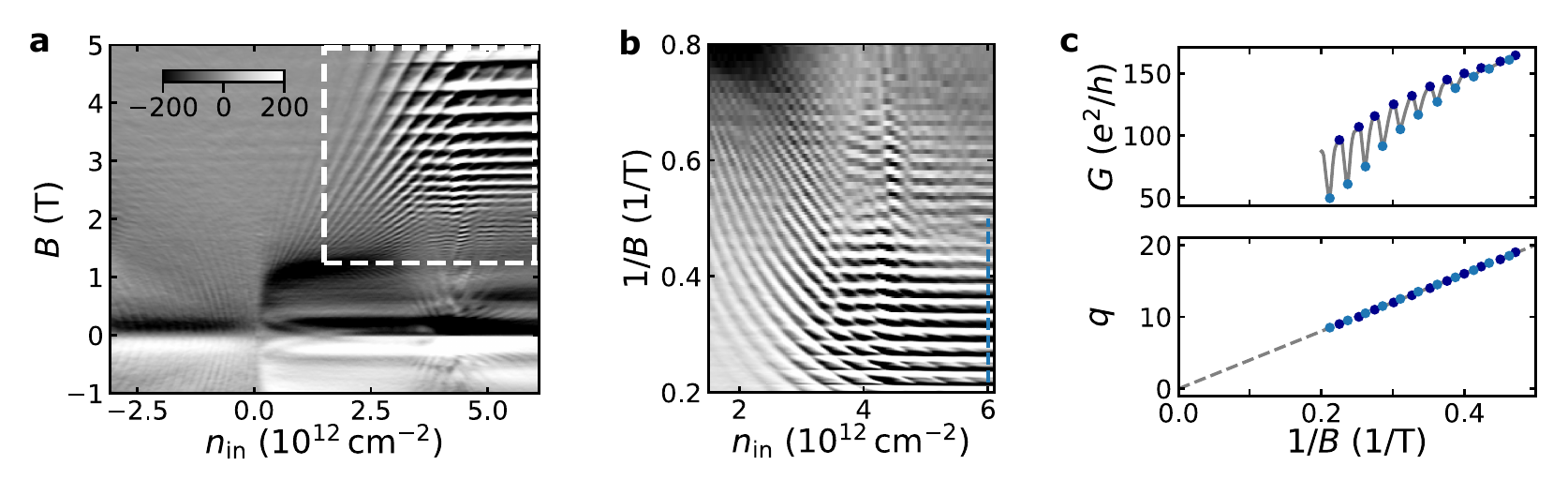}
	\caption{\setstretch{1.1}(a)~Numerical derivative of the conductance $\mathrm{d}G/\mathrm{d}B$ (in $e^2/(h\,\mathrm{V}$)) as a function of charge carrier density $n_\mathrm{in}$ and magnetic field $B$ at a constant outer density $n_\mathrm{out}=4.1\times10^{12}\,$cm$^{-2}$ (see \ref{fig:S2}a). (b)~Zoom-in on the upper right corner of panel a (white dashed box) but here plotted as a function of $1/B$. (c)~Top: Respective conductance curve $G$ as a function of $1/B$ at constant $n_\mathrm{in}=6\times10^{12}\,$cm$^{-2}$ (blue dashed trace in panel (b)). Bottom: Corresponding index $q$ of the oscillation maxima (dark blue; integer numbered) and minima (light blue; half-integer numbered) as a function of $1/B$. The gray dashed line is a linear fit.}
	\label{fig:S3}
\end{figure*}

The moir\'{e} wavelength of our superlattice structure is estimated from the Brown-Zak oscillations as shown in \ref{fig:S3}. These oscillations are density independent and arise at fractional magnetic flux through the superlattice unit cell $BA_0 = \Phi_0/q$, where $B$ is the magnetic field, $A_0$ is the superlattice unit cell, $\Phi_0$ is the magnetic flux quantum and $q$ is an integer number \cite{KrishnaKumar17,Chen17}. \ref{fig:S3}c (bottom panel) shows the oscillation index $q$ as a function of $1/B$. From the slope of the linear fit (gray dashed line) $\Phi_0/A_0=40.1\,$T, we find the moir\'{e} wavelength $\lambda\approx10.9\,$nm. A similar value $10.4\,$nm was used for the electrostatic superlattice potential of the quantum transport simulations, obtained as the best match for the position of the satellite Dirac points in the density-density-maps (cf.~Figs.~2a,\,b in the main text).

\section{Source-drain bias spectroscopy of Fabry-P\'{e}rot interferences}

\begin{figure}
	\centering
	\includegraphics{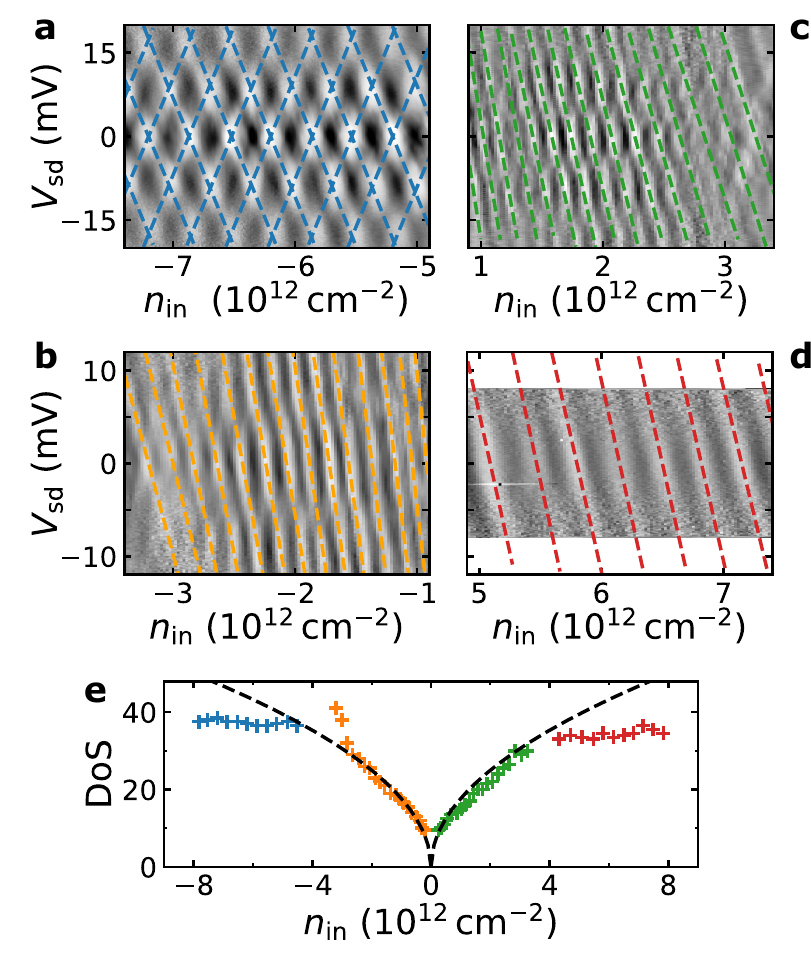}
	\vspace{-0.3cm}
	\caption{(a-d)~Source-drain bias spectroscopy of Fabry-P\'{e}rot interferences of the Fabry-P\'{e}rot interferences in cavities pPp, npn, pnp and nNn, respectively, showing the conductance oscillations as a function of charge carrier density $n_\mathrm{in}$ and source-drain bias voltage $V_\mathrm{sd}$ at constant outer charge carrier densities $n_\mathrm{out}=-0.7\times10^{12}\,$cm$^{-2}$, $2.0\times10^{12}\,$cm$^{-2}$, $-2.0\times10^{12}\,$cm$^{-2}$ and $1.4\times10^{12}\,$cm$^{-2}$, respectively. (e)~Extracted density of states (DoS) in units of $10^{12}\,$eV$^{-1}\,$cm$^{-2}$ as a function of charge carrier density $n_\mathrm{in}$ for all four cavities. Black dashed lines correspond to a theoretical $\mathrm{DoS}\propto\sqrt{n}$ of massless Dirac fermions.}
	\label{fig:S4}
\end{figure}

It is possible to utilize source-drain bias spectroscopy of FP interferences to probe the density of states (DoS) of the superlattice minibands within the cavity \cite{Cho11}. The four junction configurations pPp, npn, pnp and nNn are shown in \ref{fig:S4}a-d, respectively, as a function of charge carrier density and applied source-drain bias voltage $V_\mathrm{sd}$. As one can see, the obtained checkerboard patterns depend on the gate conditions, \textit{i.e.} the transmissions across the barriers which form the cavities \cite{Wu07,Pandey19}. By assuming the slope $\Delta V_\mathrm{sd} /\Delta n_\mathrm{in}$ of the linearly shifted resonances to be inversely proportional to the density of states $\mathrm{DoS}\equiv\mathrm{d}n/\mathrm{d}E$, where the change in energy is determined as $\Delta E=e\Delta V_\mathrm{sd}/2$, an approximate measure of the DoS can be extracted \cite{Cho11}. \ref{fig:S4}e shows the obtained values together with a theoretically expected square-root-dependence $\mathrm{DoS} \propto \sqrt{n}$ (black dashed lines) for the case of massless Dirac fermions in monolayer graphene given by $2\sqrt{(n/\pi)}/\hbar v_\mathrm{F}$, where $v_\mathrm{F}=3ta_\mathrm{CC}/2\hbar$ is the Fermi velocity with $t=3\,$eV the tight-binding hopping parameter and $a_\mathrm{CC}=0.142\,$nm the carbon--carbon bond length. In regions npn and pnp the estimated DoS is in good agreement with the theoretical curve. We further note, that the extracted DoS features a pronounced increase when approaching the secondary Dirac point on the hole-side, which could be a signature of the van Hove singularities in the vicinity of the satellite Dirac points \cite{Indolese18}. Yet, the patterns become less clear within this range. Going beyond the secondary Dirac points where the observed FP interferences arise due to confinement of superlattice charge carriers, the DoS shows a sudden and drastic different behavior, indicative of massive particles in the superlattice minibands. However, it should be noted that the employed two-terminal measurements also include the contact resistance, which is neglected here and the voltage drop is assumed to only occur across the interferometer cavity itself. But even though a quantitative analysis is not possible for this reason, it becomes unambiguously clear that the observed patterns are distinct to the massless Dirac fermions case.

\section{Fabry-P\'{e}rot interference dispersion at low magnetic field}

\begin{figure*}
	\centering
	\includegraphics{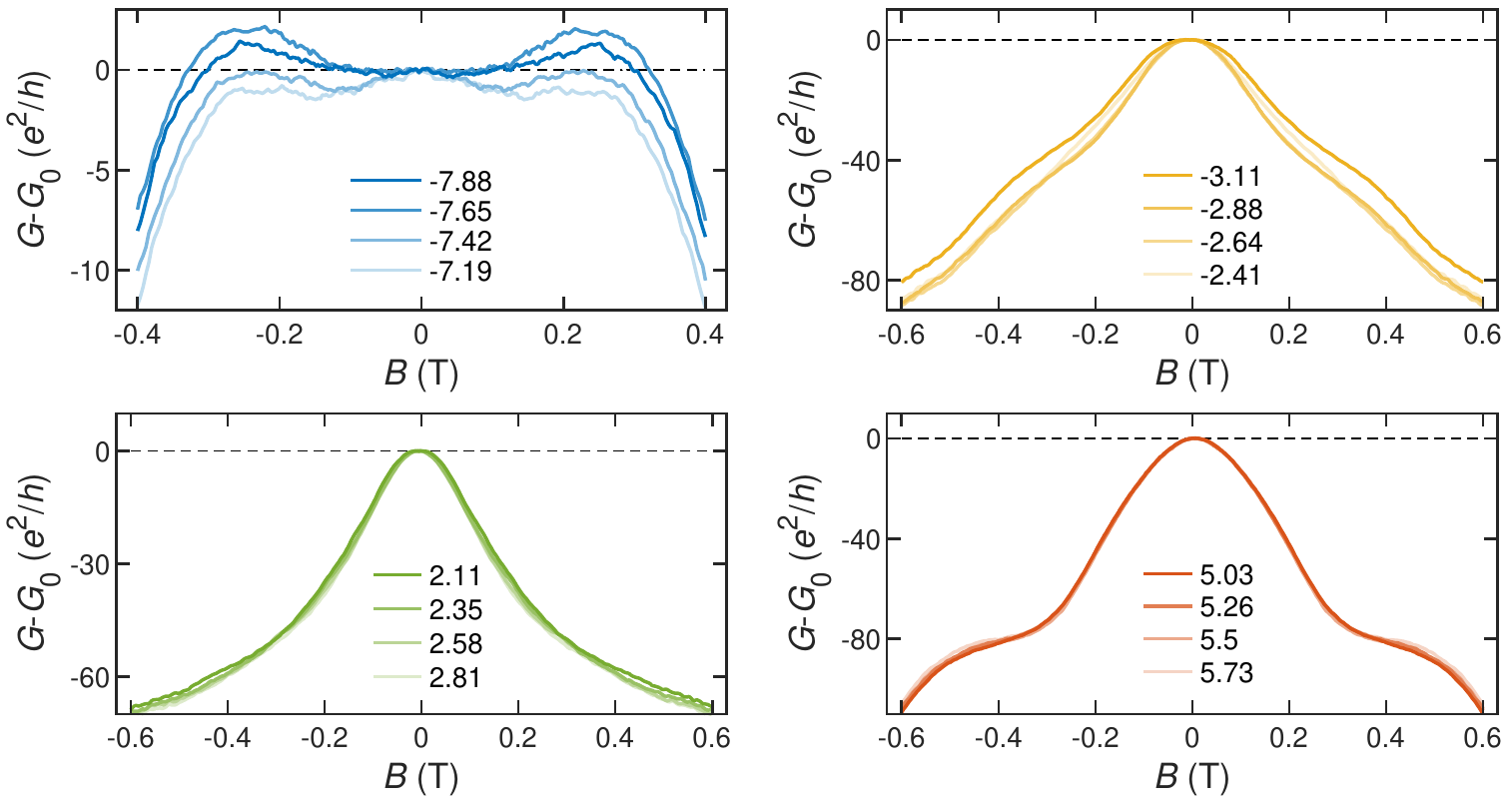}
	\caption{\setstretch{1.1} Low magnetic field measurements of the four different cavities pPp, npn, pnp and nNn at constant outer charge carrier densities $n_\mathrm{out}=-1.4\times10^{12}\,$cm$^{-2}$ (up left), $2.0\times10^{12}\,$cm$^{-2}$ (up right), $-2.0\times10^{12}\,$cm$^{-2}$ (down left) and $1.4\times10^{12}\,$cm$^{-2}$ (down right), respectively. Curves show the change in the conductance $G(B)-G(B=0)$ as a function of magnetic field $B$ for different charge carrier densities $n_\mathrm{in}$ (values given in units of $10^{12}\,$cm$^{-2}$).}
	\label{fig:S8}
\end{figure*}

\begin{figure*}
	\centering
	\includegraphics{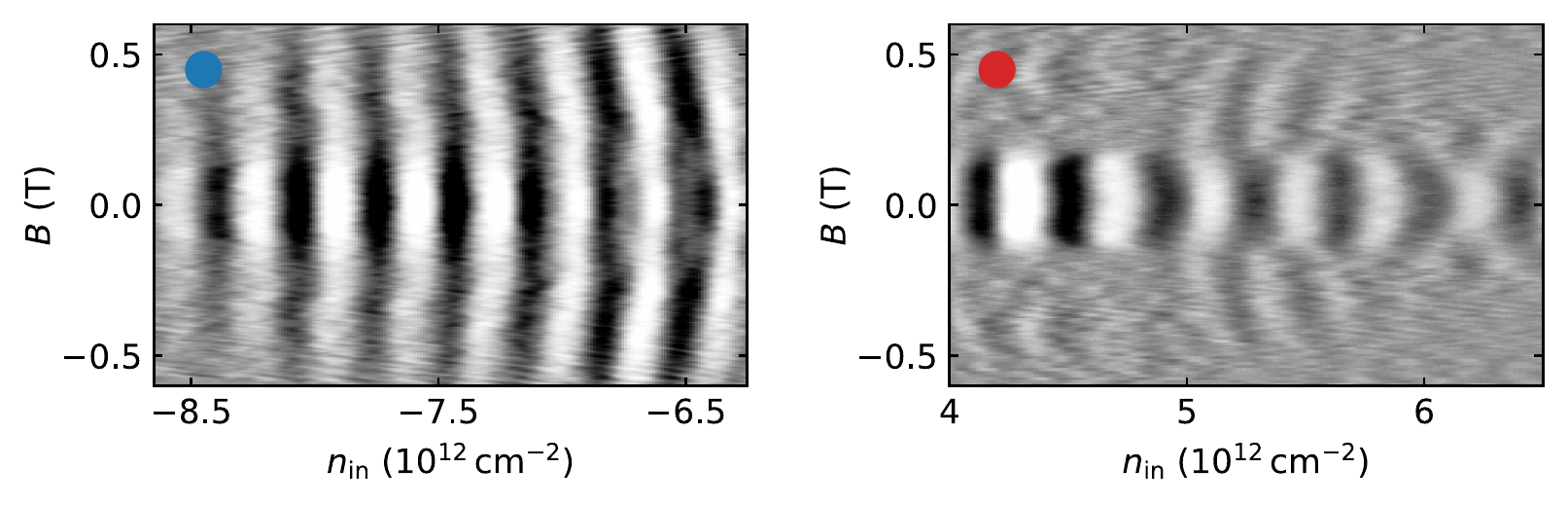}
	\caption{\setstretch{1.1} Additional magnetic field dependence data of the Fabry-P\'{e}rot interferences.~Conductance oscillations $\Delta G$ as a function of inner charge carrier density $n_\mathrm{in}$ and magnetic field $B$ in cavities pPp (left; blue) and nNn (right; red) (cf.~Fig.~3a in the main text, but here at different outer charge carrier densities $n_\mathrm{out}=-2.0\times10^{12}\,$cm$^{-2}$, and $0.7\times 10^{12}\,$cm$^{-2}$, respectively.)}
	\label{fig:S9}
\end{figure*}

\ref{fig:S8}a shows the change in the conductance as a function of magnetic field for different $n_\mathrm{in}$ at constant $n_\mathrm{out}$\,. The presented curves of regions npn and pnp follow the expectation of continuously reduced transparency upon applied magnetic field, which is due to the bending of trajectories and consequently larger incident angles on the barriers \cite{Shytov08,Young09,Katsnelsonbook}. In contrast, pPp and nNn exhibit clearly different and more subtle behaviors as seen in Fig.3 of the main text and \ref{fig:S9}. In the pPp case, we observe a widely constant or even increased conductance at finite magnetic field before it drops, whereas in nNn the conductance features an unusual plateau-like shoulder at finite magnetic field. Noteworthy, the described anomalies in the conductance appear at the same magnetic field values where the FP interferences abruptly vanish, as discussed in the main text.

\section{Quantum transport simulations}

Throughout this work, all simulations are obtained from real-space Green's function method based on a graphene lattice up-scaled by a factor of $s_f$ ($s_f=4$ for Figs.~S2b and 2b of the main text, which cover either high magnetic field or high density ranges, and $s_f=6$ for Figs.~3c,\,d of the main text, which are restricted to low magnetic field and reasonable density ranges), taking into account the scalar superlattice potential modeling the moir\'e pattern \cite{Yankowitz12}, as described in Ref.~\onlinecite{Chen19}. Local current densities reported in Fig.~3d of the main text are imaged by applying the Keldysh-Green's function method in the linear response regime \cite{Cresti03}. At each lattice site $n$, the bond charge current density $\mathbf{J}_n=\sum_m\mathbf{e}_{n\rightarrow m}\langle J_{n\rightarrow m}\rangle$ is computed, where the sum runs over all the sites $m$ nearest to $n$, $\mathbf{e}_{n\rightarrow m}$ is the unit vector pointing from $n$ to $m$, and $\langle J_{n\rightarrow m}\rangle$ is the quantum statistical average of the bond charge current operator $J_{n\rightarrow m}$ \cite{Nikolic06}. After computing for each site, the position-dependent current density profile $\mathbf{J}(x,y)=[J_x(x,y),J_y(x,y)]$ is obtained, and those reported in Fig.~3d of the main text are the magnitude $J(x,y)=\sqrt{J_x^2(x,y)+J_y^2(x,y)}$ (see also \cite{liu17}).

\begin{figure*}
\centering
\includegraphics[width=\textwidth]{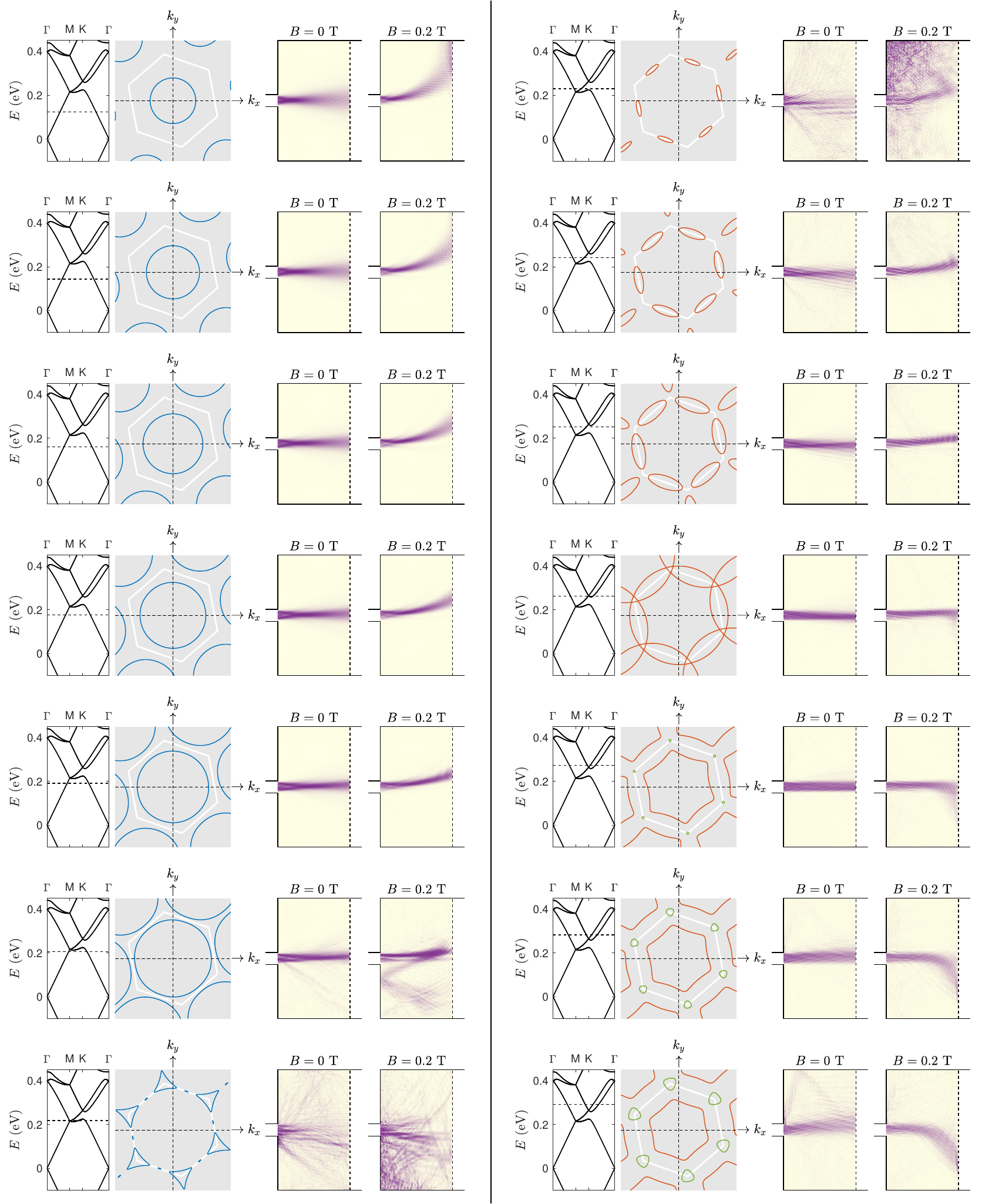}
\caption{\setstretch{1.1} Compilation of the relation between Fermi surface and charge carrier trajectories in magnetic field obtained by ``beam shooting'' simulations as described in the quantum simulations section of this Supplemental Material.}
\label{fig:S10}
\end{figure*}

\begin{figure*}
\centering
\includegraphics[width=\textwidth]{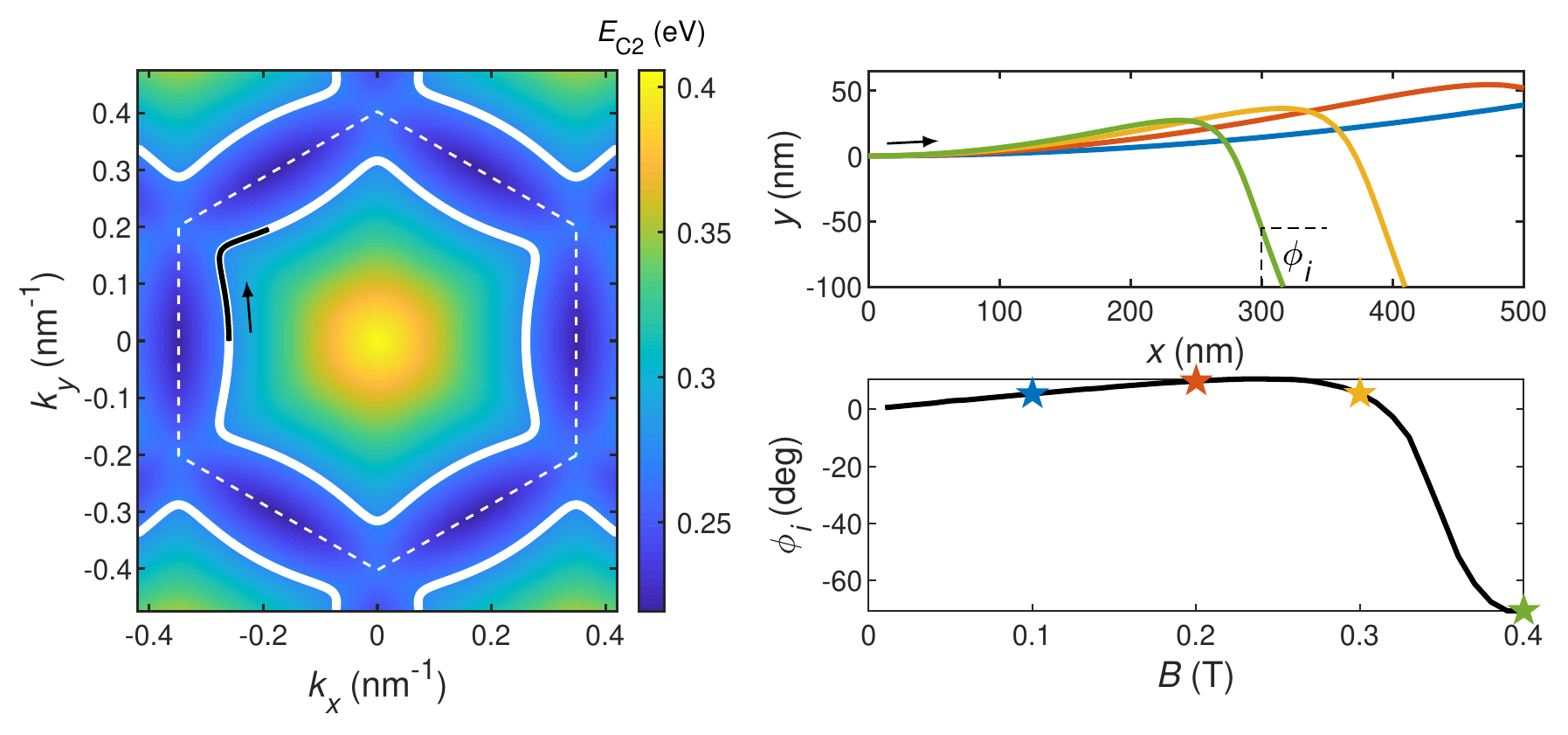}
\caption{\setstretch{1.1} Incident angle transition. ~Left: Colour map of the C$_2$ band structure $E_\mathrm{C2}(k_x,\,k_y)$. The white dashed hexagon denotes the moir\'{e} superlattice mini Brillouin zone. Additionally, the Fermi surface contour at an exemplary energy $0.2728\,$eV (same as in Fig.~3 in the main text) is shown by the white thicker lines. For simplicity, the small rotation of the superlattice structure with respect to the device orientation (see main text) is neglected here. Right upper panel: Calculated trajectories with real space coordinate $y$ as a function of position $x$ for different magnetic field values $B=100\,$mT (blue), $200\,$mT (red), $300\,$mT (orange) and $400\,$mT (green), obtained by using the semiclassical equation $\hbar\mathbf{\dot{k}}=q(\mathbf{E}+\mathbf{\dot{r}}\times\mathbf{B})$ \cite{AshcroftMerminbook} yielding a $90\,^\circ$ rotated cyclotron orbit compared to the orbit in momentum space as indicated by the small black arrow in the left panel (also see main text). Right lower panel: Respective incident angles $\phi_i$ of the trajectories onto an interface positioned at $x=300\,$nm (about the size of the interferometer cavity) as a function of magnetic field. Due to the hexagonal bending of the cyclotron motion rather abruptly changed large incident angles are noticed for magnetic fields $B\gtrsim300\,$mT corresponding to the vanishing of the Fabry-P\'{e}rot interferences. The ``star'' markers denote the respective magnetic field of the curves shown in the upper panel.}
\label{fig:S11}
\end{figure*}

To clearly reveal the electron trajectories within the same quantum transport regime and at the same time directly connect the local current density profiles to our conductance simulations reported in Fig.~3c of the main text, we consider the same sample width of 1\,$\mu$m attached to a wide drain lead (also 1\,$\mu$m in width) at the right but a thin source lead at the left (100\,nm in width). Both leads are standard graphene ribbons oriented along zigzag in the transport direction, and the moir\'e model potential is considered only in the central scattering region. Furthermore, the Fermi energy in the thin source lead is fixed at a low energy such that only the lowest mode with zero transverse momentum is injected. On the other hand, the Fermi energy in the wide drain lead is set to float with the attached right edge of the scattering region, in order to minimize the reflection that would blur the obtained electron beam profile. Additional beam simulations are shown in \ref{fig:S10}. \ref{fig:S11} displays the C$_2$ band and an example of the Fermi surface contour at an energy of 0.2728\,eV as well as the calculated real space trajectories for different magnetic fields, resulting in a drastic change of the incident angle of miniband fermion trajectories onto the barrier at a critical field.

Finally, it is important to note that the experimental findings are well captured by our quantum transport simulations using a mere electrostatic superlattice potential \cite{Chen19} (but neglecting higher order terms of the moir\'{e} perturbation \cite{Wallbank15,Moon14}). Our results are thus applicable to graphene miniband fermions subject to a hexagonal superlattice potential in general, such as recently demonstrated electrostatically induced superstructures in graphene \cite{Forsythe18,Drienovsky18}. 

\newpage

